\documentclass[10pt,preprint2]{aastex}

\shorttitle{The Dust in Lyman Break Galaxies}
\shortauthors{Vijh, Witt \& Gordon}

\begin{document}

\title{The Dust in Lyman Break Galaxies}
\author{Uma P. Vijh}
\affil{Ritter Astrophysical Research Center, University of Toledo,Toledo, OH 43606}
\email{uvijh@physics.utoledo.edu}
\author{Adolf N. Witt}
\affil{Ritter Astrophysical Research Center, University of Toledo,Toledo, OH 43606}
\email{awitt@dusty.astro.utoledo.edu}
\and
\author{Karl D. Gordon}
\affil{Steward Observatory, University of Arizona, Tucson, AZ 85721}
\email{kgordon@dirty.as.arizona.edu}

\begin{abstract}
We present our analysis of UV attenuation by internal dust of a large sample (N=906 galaxies) of Lyman Break Galaxies (LBGs). Using spectral energy distributions (SEDs) from the P\'EGASE galaxy spectral evolution model we apply dust attenuation corrections to the $G-\mathcal{R}$ colors using the Witt \& Gordon (2000) models for radiative transfer in dusty galactic environments to arrive at the UV attenuation factors. We show that the dust in the LBGs exhibit SMC-like characteristics rather than MW-like, and that the dust geometry in these systems is most likely to be represented by a clumpy shell configuration. We show that the attenuation factor exhibits a pronounced dependence on the luminosity of the LBG, $a_{1600}\propto (L/L_\sun)^\alpha$, where $0.5\leq\alpha\leq1.5$. The exponent $\alpha$ depends on the initial parameters of the stellar population chosen to model the galaxies and the dust properties. We find that the luminosity weighted average attenuation factor is likely to be in the range from $5.7-18.5$, which is consistent with the upper limits to the star formation rate at $2<z<4$ set by the FIR background. This implies that the current UV/optical surveys do detect the bulk of the star formation during the epoch $2<z<4$, but require substantial correction for internal dust attenuation.
\end{abstract}


\section{INTRODUCTION}
In the last few years, large samples of star forming galaxies have been observed at high redshift ($z > 2$) using the Lyman Break technique \citep{ste93}. So far these surveys have been carried out mostly at optical wavelengths, which explores the rest-frame UV/FUV region of the spectral energy distribution (SED) of galaxies in the range $2<z<4$. The UV luminosity in these star-forming galaxies relates directly to the number of short-lived, high-mass stars and this gives, in principle, a measure of the actual star formation rate (SFR).

Attenuation due to internal dust presents one of the biggest obstacles to interpreting these observations. Attenuation depends not only on the amount of dust, its composition and size distribution but also on its scattering properties and the geometry of the dust and its distribution relative to the sources \citep{witt92}, none of which are well known for these high-$z$ objects. The question of dust attenuation, therefore, becomes important when estimating the star formation rate history (SFRH) of the universe from observation of rest-frame UV fluxes of star-forming galaxies at high redshift. 

Many attempts have been made to estimate the SFRH in the recent years. One of the first was by \citet{mad96}. This study, however, underestimated the effects of dust. The fact that high redshift galaxies are intrinsically very dusty became apparent with the advent of the ISO observations in the mid- and far-IR \citep{pug96}. Deep surveys in the FIR gave rise to galaxy counts, and COBE data on the FIR background resulted in upper limits for the SFR \citep{elb99,pug99,alt99,aus99}. Even after correction for dust attenuation there still remained a discrepancy in the SFRs as determined by UV/optical surveys and the FIR background. There is a debate whether all the star-formation in the universe is in fact being detected by current UV/optical surveys. \citet{p&w00}, \citet{pog01} and \citet{rig00} report independent evidence on both local and high-$z$ luminous starbursts in which $\sim 70\% - 80\%$ of the bolometric flux from young stars is completly obscured by dust and remains hidden in the UV/optical surveys (even after correction for dust). \citet{a&s00} on the other hand suggest that the observed 850 $\mu m$ galaxy counts and the background can be explained by the LBG population by applying a proportionality correction to the optical flux, using the locally observed distribution of the mm-to-optical flux ratios. 

We believe these conflicts can be resolved by an improved knowledge of the dust, its properties and geometry in the high-$z$ galaxies. We present our analysis for the attenuation correction for UV emission from a large sample (N=906) of Lyman Break Galaxies (LBGs), due to internal dust, at redshift $2 < z < 4$.  This is the largest sample of spectroscopically identified high-$z$ galaxies to date. We rely on radiative transfer calculations to estimate the dust attenuation in each individual galaxy. In this approach no assumptions are made to relate different SFR indicators, nor do we apply relations which have only been established in the local universe, to high-$z$ objects. The only assumptions made are that these systems have a shell geometry and that the dust is clumpy, for both of which there are strong indications \citep{gcw97}. We explore the dependence of dust opacity on the intrinsic UV luminosity of the LBGs. We further investigate the possible evolution of the dust opacity in LBGs with respect to redshift and also the origin and nature of dust in the early universe.

The layout of the paper is as follows. In section 2 we describe the data and their limitations. Section 3 discusses the SEDs and the initial parameters used to model the intrinsic colors of the LBG population. In section 4 we describe the dust models and the attenuation functions used. Section 5 describes the analysis and the results obtained. In section 6 we present a detailed discussion comparing and contrasting our results with previous work as well as the implications of our results. Finally, section 7 summarizes our conclusions.

\section{DATA} 
Our data consisted of $\mathcal{R}$-magnitudes, colors ($U_n - G, G - \mathcal{R}$) and spectroscopic redshifts of $\sim 935$ LBGs (Charles Steidel 2001, private communication). The UV-dropout technique was used to identify these galaxies at $z\sim3$ and spectroscopic redshifts were obtained using the Low-resolution Imaging Spectrograph on the Keck telescopes. A detailed description of the filters and the observations can be found in \citet{ste93} and \citet{ste95}. For rest-frame spectra of some of these objects see \citet{pet01}. Since the U-dropout was a selection criterion for the LBGs, the $U_n - G$ color, at best, was only a 1 sigma detection. Therefore we only used the $G - \mathcal{R}$ color for our analysis. We also excluded the objects spectroscopically identified as AGNs and QSOs, reducing the sample size to 906 galaxies.

It was unfortunate that we were able to salvage only one color for such a large sample. Determining the stellar content, dust content, dust properties and the dust-star geometry, all from just one color is dangerous. Therefore, as a check for our model parameters, we applied our models to a smaller LBG sample for which longer colors were also available. We used the $V$ and $H$ magnitudes from \citet{pap01} ($31$ galaxies) and \citet{s&y98} ($17$ galaxies).  The $H$ denotes the AB magnitudes from the NICMOS F160W filter and IRIM on the Kitt Peak 4.0 m. $V$ denotes the AB magnitudes from the HST F606W filter.  The HST images were smoothed to match the much poorer PSF of the ground based IR observations. See \citet{pap01} and \citet{s&y98} for complete descriptions of the data.

\begin{figure}[h]
\plotone{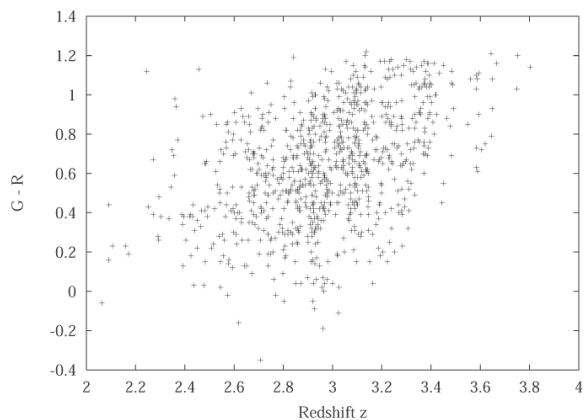}
\caption{$G-\mathcal{R}$ color plotted against spectroscopic redshift $z$ for $906$ LBGs. The upper cut-off at $G-\mathcal{R}\sim1.2$ is a selection limit. Progressive reddening at $z>3$ is because of IGM absorption \citep{mad95}.\label{figG-Rvsz}}
\end{figure}

The $G-\mathcal{R}$ sample is subject to certain selection effects: only galaxies with $G-\mathcal{R}< 1.2$ were spectroscopically studied and included in this sample, shown in Figure~\ref{figG-Rvsz}. This selection effect has significant consequences for the reddening estimates, particularly at large redshifts. No effects are introduced on the blue-limit of the observed colors, and the progressive reddening of the entire sample at $z>3$ can be attributed to absorption by the intergalactic medium. \citep{mad95}.

\section{\label{secSED}SPECTRAL ENERGY DISTRIBUTIONS}
The $G - \mathcal{R}$ color of sources at $z\sim3$ provides a measure of the FUV spectral slope in their rest frame.  The large range of $G - \mathcal{R}$ colors in the LBGs at a given redshift suggests that the sources are reddened by different amounts and that this reddening is mainly due to dust intrinsic to the sources. It is therefore necessary to obtain an unreddened spectral energy distribution (SED) from dust-free models of young star-forming populations for use as a template. We used the P\'EGASE2 galaxy spectral evolution model \citep{fio99} to model the LBG SEDs. The prominent feature of this model is that it includes stellar evolutionary tracks with non-solar metallicities and contribution from nebular emission. \citet{zac01} provide a comparison of different galaxy spectral evolution models used in recent studies. They show that the broad-band UV colors predicted by their model and the P\'EGASE2 codes are not significantly different for a large range of metallicities. There are differences in the V$-$K color, but this color is of little consequence in star-forming regions. After choosing the initial parameters to describe an LBG, the resulting SED was redshifted by the measured redshift and convolved with the $G$ and $\mathcal{R}$ transmission curves (Charles Steidel 2001, private communication; also see \citet{ste93}) to get the $G-\mathcal{R}$ colors, using codes that are a part of the P\'EGASE2 package.

The synthetic spectra are computed at specified ages after the onset of the star formation episode. The properties of the stellar populations are computed based on the shape of the initial mass function (IMF) and its upper and lower limits. Other parameters used to compute the SED are the initial metallicity of the ISM and the star formation history.  

\begin{figure}[h]
\plotone{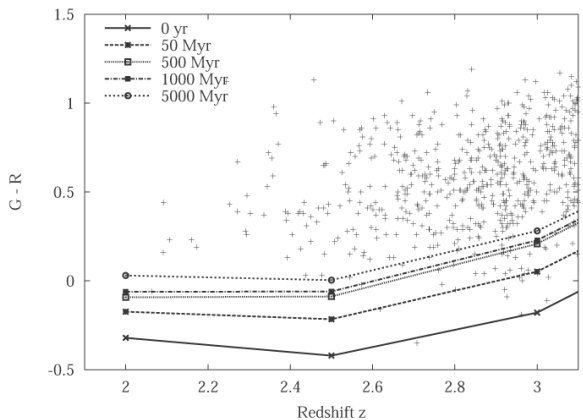}
\caption{$G-\mathcal{R}$ color predicted by increasingly older SEDs as a function of redshift $z$. The range on observed colors of the LBGs cannot be explained by age alone.\label{figage}}
\end{figure}

Working with only one UV color, we face the dilemma produced by the age-reddening degeneracy. For a given source, we are unable to distinguish between  the cases of an older (and therefore redder), dust-free population and that of a younger, dust-reddened population. There are various factors that can redden the colors of a galaxy; age of the stellar population, increased metallicity, higher amounts of dust or any combination of these factors. In principle, this degeneracy cannot be broken working with one UV color alone. In our analysis we break this degeneracy arbitrarily by analyzing the sample under two sets of assumptions regarding the time dependence of dust formation in these high-z galaxies, as discussed in the next two subsections. In each of the two cases we assume a fixed age of the stellar population for all the LBGs and assume conditions that leads to an upper-limit and a lower limit on the dust attenuation. We find that only a small fraction of the color range of the LBGs can be attributed to age alone; most of it must be due to reddening by internal dust. As shown in Figure~\ref{figage}, at a given redshift ($z\sim2$), the $G-\mathcal{R}$ color gets redder by $0.26$ as the galaxy ages from $0\ yr$ to $1\ Gyr$ and by $0.15$ as the galaxy ages from $50\ Myr$ to $1\ Gyr$. The color range for the observed LBGs at the same redshift is $1.6$. Analysis under two different sets of assumptions based on the timescale of dust formation helps in resolving this age-reddening degeneracy as far as it exists. Furthermore we verify our model parameter assumptions by applying them to a subset of the LBGs for which longer colors ($V-H$) were available.  At $z\geq3$, the colors of a galaxy are further reddened by IGM absorption. Both optically thin Lyman$-\alpha$ forest clouds and optically thick Lyman-limit systems contribute to the reddening of background sources \citep{mad95}. We have corrected our assumed SEDs for this effect using the values provided in \citet{mad95}.

\subsection{\label{secSEDI}Case I}

In this case we assumed that we are observing the LBGs within a short time into their first episode of star formation. The dust free model was chosen to represent the very youngest, bluest LBG. This implies that only the very bluest of the LBGs are dust free and that increased reddening is a consequence of rapidly forming dust. We further assumed that dust formation follows the formation of massive stars on a nuclear time scale. In order to get an absolute lower envelope to the data at higher redshifts ($z>2.5$) we had to assume lower metallicities (and correspondingly younger ages). The initial metallicity of the ISM was varied from $0.1\ Z_\sun$ to $0.01\ Z_\sun$ with $z =2$ to $z=4$ respectively. In order to mimic the behavior of the data at the blue envelope we had to increase the higher mass limit for the IMF at larger redshifts.  We used a Salpeter initial mass function with a lower mass limit of $0.1\ M_\sun$. Studies of the early metal pollution of the intergalactic medium show that the IMFs are increasingly biased toward higher masses at large redshifts (see \citet{abia01} and references therein). Also \citet{lfr02} show that if massive stars are the progenitors of long lived gamma ray bursts, the IMFs at large redshifts need to be flatter with much higher upper mass limits. The upper mass limit for our models was varied from $120\ M_\sun$ to $200\ M_\sun$ with $z=2$ to $z=4$ respectively. We divided the sample into four redshift bins ($2 \leq z < 2.5,\ 2.5 \leq z < 3,\ 3 \leq z < 3.5,\ \textrm{and}\ 3.5 \leq z < 4$), to simulate the changing behavior of the data with redshift as well as to study the possible evolution of the dust properties. Table~\ref{case1&2prop} shows the parameters  which were  required to match the bluest observed colors with an unreddened SED for case~I. A constant Star Formation Rate (SFR) of $100\ M_\sun\ yr^{-1}$ was used at all redshifts. The colors were examined at different ages after the star formation had started as noted in Table \ref{case1&2prop} under the column labeled ``Age''. Values for the UV-spectral slope $\beta$ were calculated as prescribed by \citet{cal01} for the spectral range $0.126\ \mu m < \lambda < 0.26\ \mu m$. A scenario favoring an IMF biased toward massive stars, low stellar metallicity, and substantial dust formation on a $10^7\ yr$ time scale is supported by the equivalent width distribution and number counts of Lyman$-\alpha$ emission galaxies in the early universe by \citet{m&r02}.  The intrinsic colors of all the galaxies are assumed to correspond to this young model.

In this case the model parameters were chosen so as to provide a blue envelope to the sample, representing the youngest unreddened galaxies. In doing this, we have assumed a short time-scale for dust formation and assumed that only the youngest systems are dust free. In such young star forming systems Type II supernovae (SNe II) are the dominant sources of dust production. (For a detailed study on the observed properties of SN dust see \citet{wod97}). Modeling the dust content of a nearby low metallicity galaxy incorporating dust from SNe II \citet{hhf02} show that the dust mass reaches its maximum in $\sim2\times10^7\ yr$. \citet{e&e98} model dust masses in galaxies and show that the dust content in galaxies with high star formation rates advances within a factor of $2$ of the maximum values at very young ages. For $m_\star/m_{tot}=0.2$, if one considers $m_{tot}=10^{11}\ M_\sun$, and SFR $=100\ M_\sun yr^{-1}$, dust masses grow within a factor of $2$ of the maximum in $2\times10^7\ yr$. This assumption of the short time scale for dust formation results in larger dust attenuation factors for these galaxies.

\subsection{\label{secSEDII}Case II}
In this case we assumed longer timescales for dust formation, consistent with the assumption that mass loss from evolved lower-mass stars is the principal contributor to dust formation. This process follows the slower ($\tau\sim10^9$ yr) evolutionary time scale of the stars; it is more significant in the present universe and less so in the high-z universe \citep{ed01}. The model parameters for this case were chosen, so as to arrive at a conservative lower limit to the amount of dust present in LBGs responsible for partial attenuation of their UV flux. \citet{sha01} analyzed 81 LBGs for which they also had K$_s$ and J band fluxes, and these were found to have ages between 10 Myr and 1 Gyr with a median value of 320 Myr. We assume that galaxies with populations younger than $320$ Myr are dust free and that dust appears only through subsequent evolution. The $320$ Myr old SED is therefore used as a reference for all LBGs with colors redder than those implied by this reference.  As we were not attempting to model the bluest LBGs, dividing the sample into different redshift bins would not constrain the model parameters in any way. Therefore, in this case we do not divide the sample into different redshift bins. A constant SFR of $100\ M_\sun\ yr^{-1}$ and a Salpeter IMF with lower limit of $0.1\ M_\sun$ and an upper limit of $125\ M_\sun$ were chosen, which are consistent with those found by \citet{sha01}. The initial metallicity of the ISM was set to $0.2\ Z_\sun$ and the colors were sampled at $320\ Myr$. This larger age and higher metallicity vis-a-vis case I result in redder colors for the dust-free (unreddened) galaxies, and all the galaxies with colors bluer than this were assumed to be dust-free. Table~\ref{case1&2prop} shows the parameters used for the analysis of this case.

For this case, we assume that the galaxies $320\ Myr$ old and younger are still dust free. Dust forms in these galaxies at later times. This long time-scale for the dust formation implies that the dust formation in these galaxies was primarily a result of the evolved intermediate mass stars and the dust production time-scales are linked to the evolution time-scales of intermediate mass stars. As explained earlier, we adopt an age of $320\ Myr$ to account for the fact that \citet{sha01} found that this was the median age of LBGs using a multi-wavelength study, and this approach results in a lower estimate to the amount of dust in these systems. 

These two cases are not just two possible scenarios, but from the viewpoint of dust formation scenarios are extreme models, and their range of results will cover any reasonable intermediate model. 

\begin{deluxetable}{lcccc}
\tablecaption{Parameters assumed for SEDs under Case I and Case II\label{case1&2prop}}
\tablecolumns{5}
\tablewidth{0pt}
\tablehead{
\colhead{Redshift Range}&\colhead{Initial Metallicity} & \colhead{Age} & \colhead{Upper mass limit} & \colhead{$\beta$}
}
\startdata
\sidehead{Case I}
$2.0<z\leq2.5$ & $0.1\ Z_\odot$ & $\sim50\ Myr$ & $120\ M_\odot$ & $-3.14\pm0.04$ \\
$2.5<z\leq3.0$ & $0.1\ Z_\odot$ & $\sim50\ Myr$ & $135\ M_\odot$ & $-3.5\pm0.02$ \\
$3.0<z\leq3.5$ & $0.01\ Z_\odot$ & $\sim0\ yr$ & $150\ M_\odot$ & $-3.76\pm0.02$ \\
$3.5<z\leq4.0$ & $0.01\ Z_\odot$ & $\sim0\ yr$ & $200\ M_\odot$ & $-3.82\pm0.02$ \\
\sidehead{Case II}
$2.0<z<4.0$ & $0.2\ Z_\sun$ & $\sim320\ Myr$ & $125\ M_\sun$&$-3.04\pm0.04$\\
\enddata
\end{deluxetable}

\section{\label{secDAf}DUST MODELS AND ATTENUATION FUNCTIONS}
To relate the SEDs generated with the assumptions outlined in Sec.~\ref{secSED} to the observed colors, we need to know the wavelength dependence of the dust attenuation function in these galaxies. It is important to distinguish between the terms extinction and attenuation. The term ``extinction law '' was originally defined for stars. It quantifies the wavelength dependence of dust absorption and scattering out of the line of sight toward point sources. In distant galaxies however the situation is quite different. There are two main reasons: (i) The complex geometry of the light sources and the dust distribution and (ii) the flux scattered off the dust grains is returned into the line of sight in complex ways. Thus the ``attenuation'' depends critically on the geometry of the dust and the sources and the optical depth and scattering properties of the dust. 

\begin{figure}
\plotone{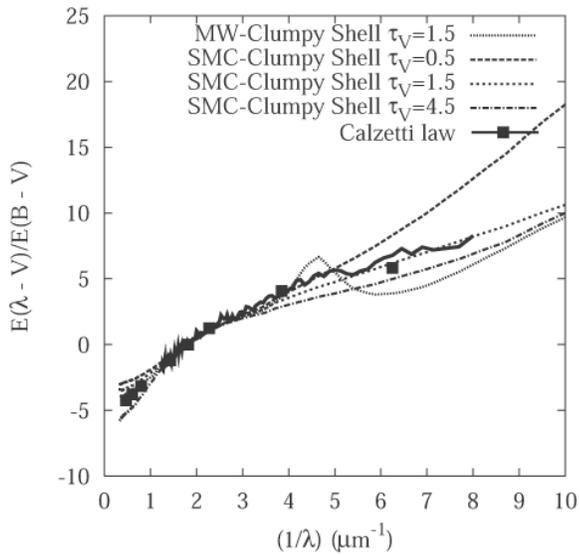}
\caption{Attenuation functions from clumpy shell models at different optical depths $\tau_V$ compared to the Calzetti law. Dotted line is the attenuation described by $E(\lambda-V)/E(B-V)$ for MW-like dust at $\tau_V=1.5$. Long-dashed line is the attenuation curve for SMC-like dust at $\tau_V=0.5$, short-dashed for SMC-like dust at $\tau_V=1.5$, and dash-dotted for SMC-like dust at $\tau_V=4.5$. Solid line represents the Calzetti law where the solid squares are photometric points. The Calzetti law can be reproduced by SMC-Clumpy shell model with $\tau_V=1.5$.\label{figaf}}
\end{figure}

In recent years the wavelength dependence of dust attenuation in star forming galaxies has been estimated by the so called ``Calzetti law'', which is an empirical determination of the observed attenuation function derived by averaging over a small sample of UV-bright local starburst galaxies \citep{cal94,cal95,cal00}. This empirical law leads to a correlation between the UV spectral slope $\beta$ and the FUV attenuation. The FUV flux has then been corrected for attenuation, assuming this correlation holds for all redshifts. This law however has limitations; while applicable to the particular sample there is no evidence that this law applies over all redshifts or that it applies to systems with substantially more dust than is present in the local sample of UV-bright galaxies. \citet{bell02} demonstrates that normal star-forming galaxies deviate substantially from the starburst $\beta-A_{FUV}$ correlation. We have therefore chosen to address the question of dust attenuation with the radiative transfer models of \citet{wg00} (henceforth WG2000). These models incorporate multiple scattering radiative transfer calculations for different galactic environments, filled with either homogeneous dust or a two-phase clumpy dust distribution.  Figure~\ref{figaf} shows the attenuation functions using the clumpy shell models for various optical depths. In these models, $\tau_V$ should be interpreted as a measure of the dust mass ($\tau_V\propto$ dust mass) where the actual amount of dust would scale with the size of the system. As the amount of dust, measured by the visual optical depth for a corresponding homogeneous model increases, the attenuation law becomes flatter in the far-UV. Thus, as Figure~\ref{figaf} shows, there is not a single attenuation law, e.g. the Calzetti ``law'', but a series of attenuation laws with a functional form dependent upon the amount of dust present. WG2000 have also shown that the Calzetti law corresponds to the attenuation function for a clumpy shell galaxy with SMC dust of intermediate dust optical depth of $\tau_V=1.5$ (see Figure~\ref{figaf}). One cannot expect all the LBGs to have the same amount of dust given their wide range of observed colors.

The clumpy models of WG2000 assume a spatial filling factor of $0.15$ for denser clumps, embedded into a low density inter-clump medium, with a density contrast of $100:1$. As shown in an earlier study \citep{wg96}, this set of conditions leads to a power-law spectrum of cloud masses that closely resembles that found in the galactic ISM. The dependence of radiative transfer on other choices of clumpiness parameters was fully explored by \citet{wg96}. In the absence of any other information, we assume that the structural details of the ISM in LBGs resemble that of the MW galaxy. The geometry of the WG2000 shell models assume an inner volume ($r\leq0.3$R) occupied by stars, surrounded by a shell ($0.3\textrm{R}<r\leq$R) of dust distributed either homogeneously or in a two-phase clumpy distribution.  While relative sizes of star-filled volume and the dusty shell are to some degree arbitrary, they are modeled after he appearance of giant star-forming complexes, e.g. NGC~604 in M33 \citep{hunt96}. Important aspects are that the dust is situated between the sources and the observer, thus maximizing the attenuation, yet close enough to the sources so that scattered light becomes an integral part of the flux received from the complex.

\begin{figure}
\plotone{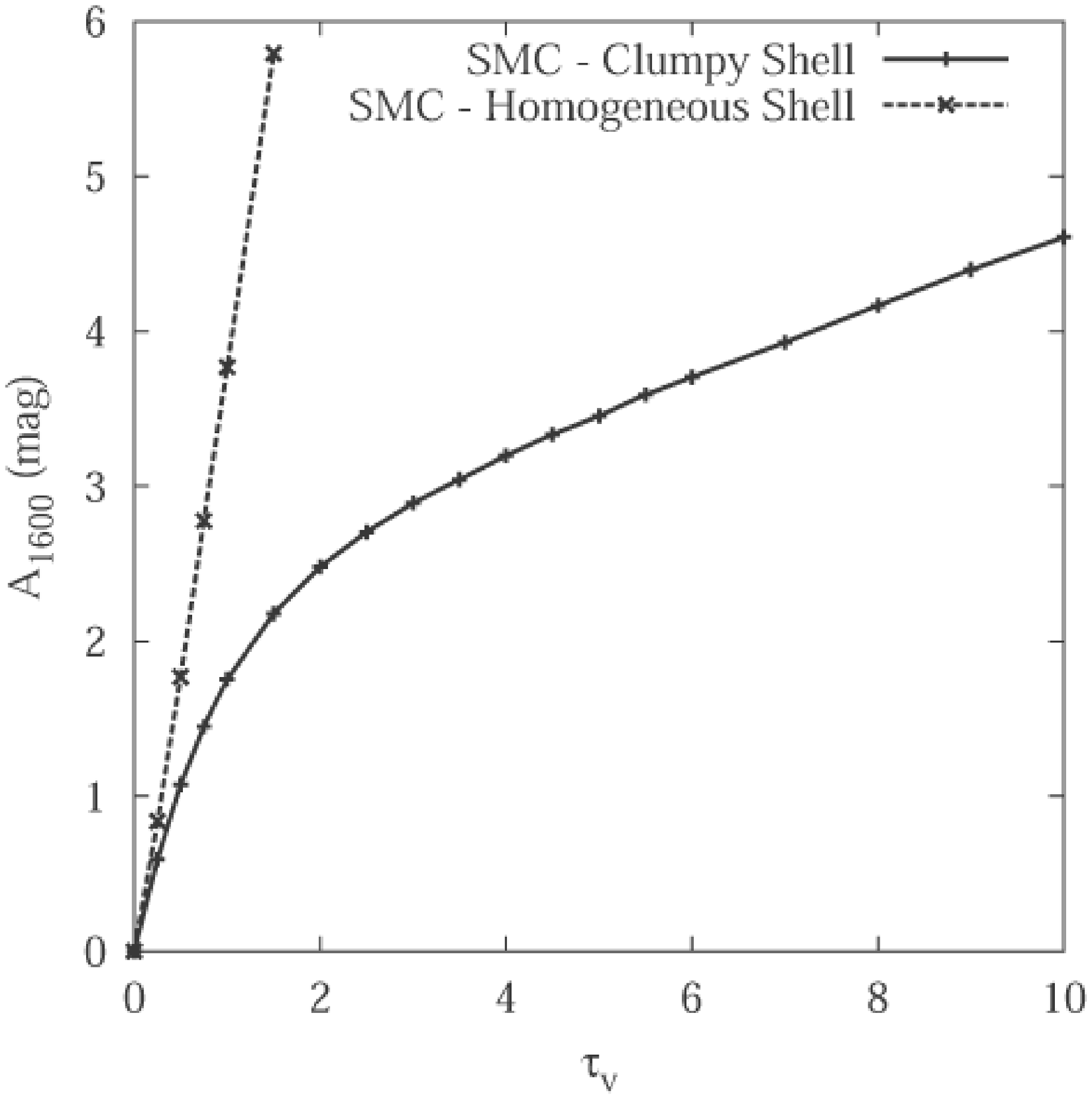}
\caption{Attenuation at $1600$ \AA$\ $, $A_{1600}$ (mag) against dust optical depth, $\tau_V$ for different geometries. Solid line represents SMC-like dust in a Clumpy-shell geometry and dashed line represents SMC-like dust homogeneously distributed in a shell geometry. \label{figAtau}}
\end{figure}

 In this study we have used the clumpy shell models to estimate the dust attenuation for the LBGs (See sec. \ref{secNG} for a more detailed analysis of the dust geometry). Figure~\ref{figAtau} shows the relationship between the attenuation at $1600$ \AA$\ $and the optical depth for SMC-clumpy shell and SMC - homogeneous shell models. In the homogeneous models the dust has uniform spatial density. This effectively represents a screen geometry, and the attenuation rises linearly with optical depth. Whereas in the two-phase clumpy models, at low values of $\tau_V$ both the dust  clumps and the inter-clump medium are optically thin and the attenuation increases proportional to the amount of dust, similar to the homogeneous case. At higher values of $\tau_V$ the clumps are starting to become optically thick but the inter-clump medium is still optically thin and the attenuation varies less rapidly with the amount of dust. At intermediate optical depths the attenuation is influenced both by the dust clumps and the inter-clump medium.

\section{ANALYSIS}
\subsection{\label{secNG}The Nature and Geometry of the Dust in LBGs}
There are significant variations in the dust properties in different galactic environments of coeval galaxies within the local group itself. In the LMC, the UV extinction curves show a distinctly different behavior between the 30 Dor region (a mini-starburst) \citep{wal91} and the rest of the LMC \citep{c&m85,fit85,fit86,mcg99}. The $2175$ \AA$\ $ bump is weaker and the far-UV rise is stronger near the 30 Dor region than in the rest of the LMC. In the bar-region of the SMC, the average extinction curve is characterized by a roughly linear rise (vs. $\lambda^{ -1}$) increasing toward shorter wavelengths without a $2175$ \AA$\ $ bump \citep{pre84,tho88,gc98}. Yet, there is one sight line that has an extinction curve with a significant $2175$ \AA$\ $ bump \citep{leq82,gc98}. In M31, the extinction curve is consistent with that of the average Galactic extinction within the associated uncertainties, although the $2175$ \AA$\ $ bump may be weak \citep{bia96}. Using IUE data for sight lines through low-density regions of the MW disk and halo \citet{cla00} show that many of the sight lines have extinction curves with weak bumps and very steep far-UV extinction reminiscent of the Magellanic Clouds. Modeling the SEDs of 30 starburst galaxies \citet{gcw97} found that the dust in these systems has extinction properties similar to the SMC lacking the $2175$ \AA$\ $ feature present in the galactic extinction curve. However, \citet{mot02} report on the detection of a significant $2175$ \AA $\ $ feature in the extinction curve of a normal-type galaxy at $z=0.83$. \citet{mas01a} report that some galaxies in the HDF spectroscopic sample show the $2175$ \AA $\ $ feature. 

\begin{figure}
\plotone{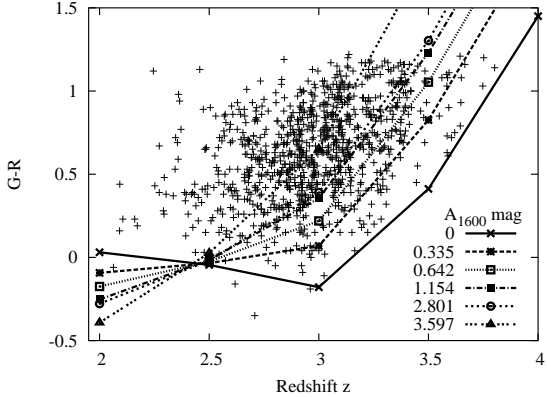}
\caption{$G-\mathcal{R}$ color vs. redshift where the different lines are the colors obtained for the intrinsic SED reddened with different amounts of MW-like dust. The attenuation (in magnitudes) produced at $1600$ \AA$\ $, $A_{1600}$  is indicated for each of the lines. The values of $\tau_V$ in the model that produce these attenuations are $0, 0.25, 0.5, 1.0, 5.0, \textrm{and } 10.0$. Even large amounts of dust cannot explain the red colors in the $2\leq z\leq3$ region.\label{figmw}}
\end{figure}

\begin{figure}
\plotone{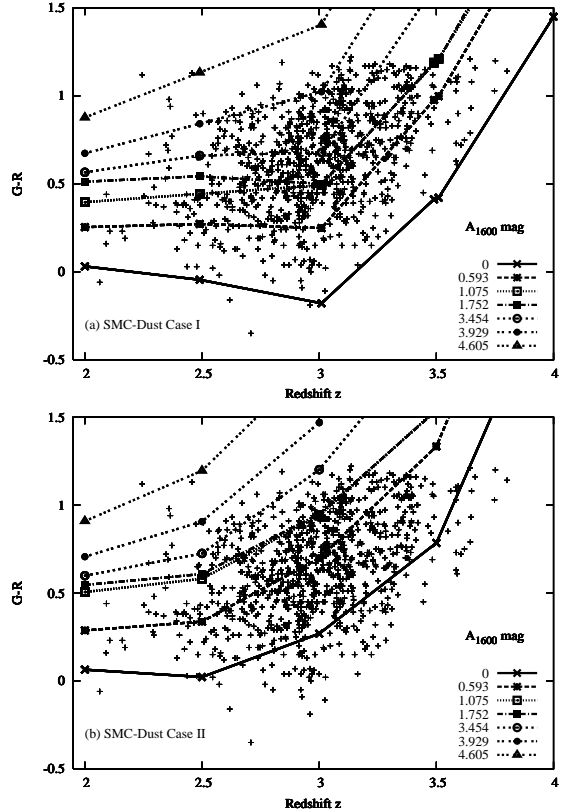}
\caption{$G-\mathcal{R}$ color vs. redshift with the colors obtained for the SED reddened with different amounts of SMC-like dust, (a) under Case~I assumptions and (b) under Case~II assumptions. The attenuation produced at $1600$ \AA$\ $ is indicated for each of the lines. The values of $\tau_V$ in the models that produce these attenuations are $0, 0.25,0.5,1.0,5.0,7.0, \textrm{and } 10.0$. \label{figsmc}}
\end{figure}

The limited evidence available so far suggests that the absence of the $2175$ \AA$\ $ feature is associated with high SFR, but this is far from conclusive. In the presence of this ongoing debate, the nature of the dust in the LBGs needs to be addressed. We use the fact that the presence of the $2175$ \AA$\ $ feature has a drastically different effect on the $G-\mathcal{R}$ color in the $2<z<2.6$ range compared to dust without the $2175$ \AA$\ $ feature to test the LBG data for the presence of the $2175$ \AA$\ $ band. We reddened the assumed SED (case~I) with the WG2000 clumpy attenuation functions with increasing optical depths of $\tau_V=0.25, 0.5, 1.0, 5.0, 7.0,\ \&\ 10.0$, with both MW- and SMC-like dust. These SEDs were then redshifted and the $G-\mathcal{R}$ color was calculated at the different redshifts. For $z\geq3$ the colors were corrected for IGM absorption using values in \citet{mad95}. Figure~\ref{figmw} is a plot of the $G - \mathcal{R}$ color vs. the redshift, where the different lines are the colors obtained for the intrinsic SED reddened with different amounts of MW-like dust. We find that MW-like dust cannot explain the range of observed colors in the $2<z<3$ range. As the $2175$ \AA$\ $ feature passes through the $G$ and $\mathcal{R}$ filter set at different $z$, the predicted $G - \mathcal{R}$ color first gets bluer (till $z<2.45$) and then gets progressively redder for increasing amounts of dust (see Figure~\ref{figmw}). On the other hand, the curves modeled by SMC-like dust (Fig. \ref{figsmc}) for increasing values of attenuation span the observed range of colors quite well. Using color-color plots \citet{gsc99} also reached the conclusion that the dust in starburst galaxies in the  North and South Hubble Deep Field is similar to SMC-like dust. Figure~\ref{figsmc}(a) shows the model-predicted curves for SMC clumpy dust under the case~I assumptions for the intrinsic SED, and Figure~\ref{figsmc}(b) for the case~II assumptions. {\sl We conclude that the dust in this sample of LBGs is better represented by SMC-like dust, which lacks the $2175$ \AA$\ $ feature.}  

\begin{figure}
\plotone{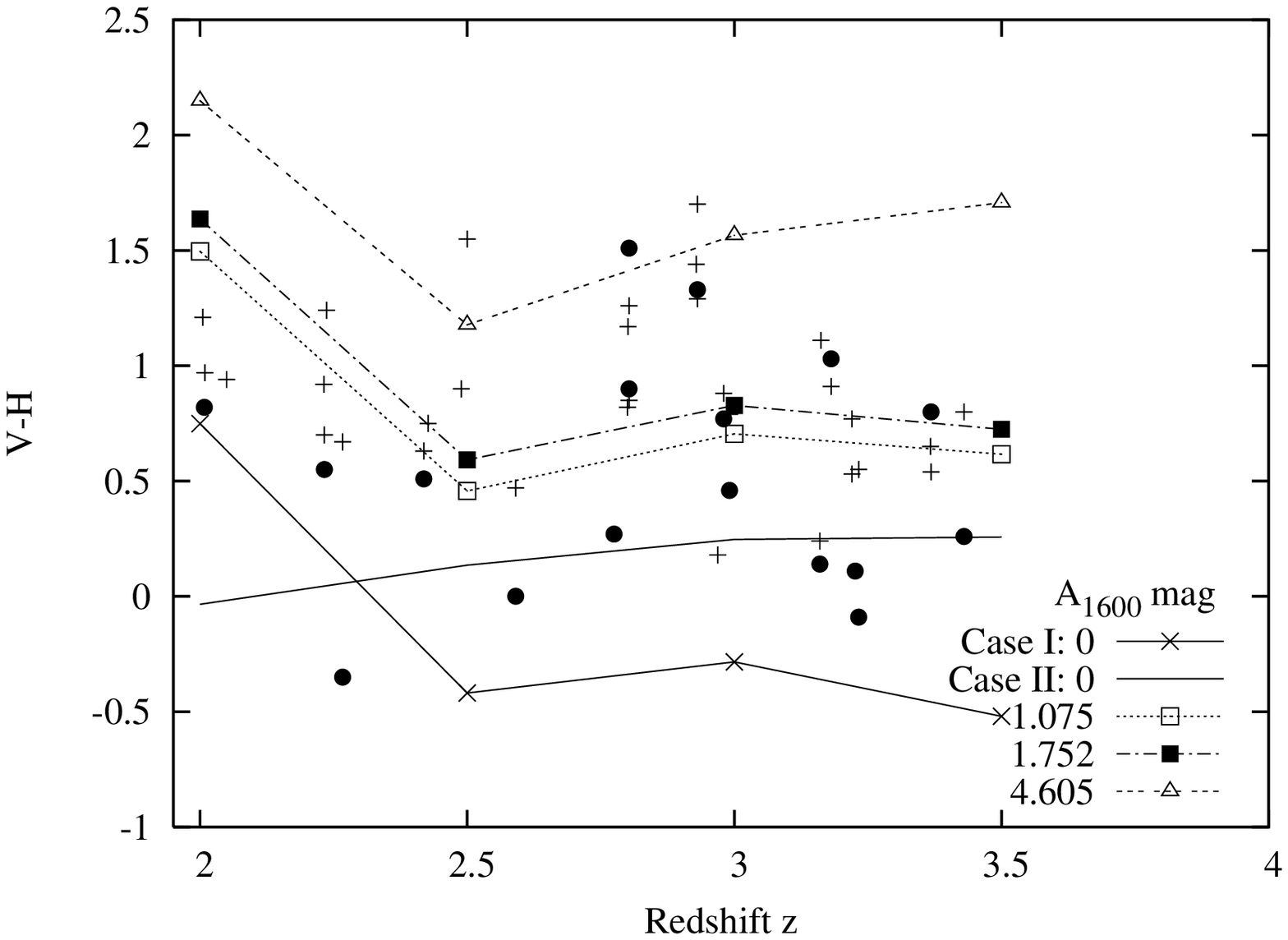}
\caption{$V-H$ color vs. redshift where the different lines are the colors obtained for an SED reddened with different amounts of SMC-like dust. The data points denoted by 'plus' symbols are from \citet{pap01} and those denoted by solid circles are from \citet{s&y98}. The line styles and symbols used are the same as those used in Figure~\ref{figmw} and Figure~\ref{figsmc}. \label{figvh}}
\end{figure}

Even though  multi-color information is available for some LBGs, such samples are much smaller and therefore less complete, being adversely affected by observational limitations. We have chosen to work with the largest sample of LBGs available to date. Unfortunately this implies that we infer dust content, dust properties, stellar content and dust-star geometry all with a single color. As a check to our model SED assumptions, we analyzed a smaller sample of LBGs for which longer wavelength colors were available. Figure~\ref{figvh} is the $V-H$ color vs. redshift for LBGs (data from \citet{pap01} and \citet{s&y98}). The curves on the figure are model-predicted curves for SMC clumpy dust, with the dust-free intrinsic SED produced  under the same assumptions as our case~I analysis. The colors predicted by a dust-free SED under case~II assumptions is also shown for reference. Comparison of Figure~\ref{figsmc}(a), \ref{figsmc}(b) with Figure~\ref{figvh} clearly shows that the smaller $V-H$ sample spans the same range in attenuations and that our model parameters are robust. This also implies that these two samples of LBGs are subsamples of galaxies in the same evolutionary stage.

We need a fundamental understanding of the environment that produces the large range of reddening that is observed. The dust geometry is a important factor in the amount of reddening produced.  For a local sample of starburst galaxies \citet{gcw97} have shown that the dust geometry is best represented by an inner dust-free sphere of stars surrounded by a star-free shell of clumpy dust such as represented by the WG2000 shell models. \citet{ino02} has found the dust in HII regions to be best described by a shell geometry. Also \citet{bua02} find that the clumpy shell geometry as described in WG2000 fits very well with their analysis of the UV SEDs of starburst galaxies. Different dust geometries are analyzed in WG2000. Not all geometries produce the same amounts of reddening for given amounts of dust; to produce equal amount of reddening other geometries would require much larger dust masses. In their Figure~11 WG2000 show that only the shell models reproduce as large a range of $\Delta \beta$ as observed. One should note that the observed $G-\mathcal{R}$ color and $\beta$ are not identical. The $G-\mathcal{R}$ color is a rest-frame UV-color, whose rest-frame UV base wavelengths vary with redshift ($G=1610$~\AA$\ $ at $z=2$ to $966$~\AA$\ $ at $z=4$, $\mathcal{R}=2310$~\AA$\ $ at $z=2$ to $1386$~\AA$\ $ at $z=4$), while the wavelengths for $\beta$ are fixed ($1200$~\AA--$2600$~\AA).

It is not surprising that a large fraction of the galaxies at $z>2$ are best described as resembling centrally concentrated ``blobs'', which are probably the progenitors of our present-day galactic bulges and elliptical galaxies \citep{van02}. The choice of a spherical star-forming region, relatively free of dust surrounded by a dust shell seems to be an appropriate geometry to describe these objects. A justification for the clumpy-shell geometry of the LBGs is due to their large SFRs, up to $\sim 1000\ M_\sun yr^{-1}$ in some cases. Such massive star formation activity will blow away the gas and dust into the surrounding region. Stars are expected to spend the very beginning of their evolution deeply embedded in dusty environments, later dispersing the molecular clouds in which they were born \citep[and references therein]{cal94}. {\sl In our analysis too, we find that the shell geometry best explains the observed colors for reasonable values of dust optical depths.}
  
\subsection{The Relation Between the Dust Opacity and the Intrinsic UV Luminosity of LBGs}
To estimate the intrinsic UV luminosity of the LBGs we need to adopt a cosmology: throughout this paper we use H$_0=65\ km\ s^{-1} \ Mpc^{-1}, \Omega_M=0.3, \Omega_\Lambda=0.7$, unless otherwise specified. The reddened SEDs were redshifted and the colors were calculated for different amounts of dust, resulting in a range of attenuations at $1600$ \AA$\ $($A_{1600}$), at different redshifts. The attenuation for each galaxy was then computed. This was done as follows: The color curves corresponding to selected values of the attenuation at $1600$~\AA$\ $, $A_{1600}$, are as shown in Figure \ref{figsmc}(a) and \ref{figsmc}(b). Dividing the sample into finite-sized ranges of color surrounding each curve partially accounts for color differences which may be attributed to a range in the intrinsic ages of the SED. The attenuation for each galaxy in the rest-frame wavelength for the $\mathcal{R}$ filter is then noted. The attenuations and $z$ are used to arrive at the absolute magnitudes, luminosity distances and the intrinsic UV luminosities. We find that the bulk of the LBGs are extremely luminous, $L\sim 10^{11}-10^{12}L_\sun$, corresponding to the LIRG and ULIRG population in the present universe.

Star forming galaxies, both in the local universe and in the early universe have been shown to exhibit a correlation between dust attenuation and their intrinsic luminosity. \citet{a&s00}, in their Figure~11, relate the dust luminosity to the total luminosity of starbursts in the local universe, at $z\sim1$ and at $z\sim3$. \citet{cal95}, \citet{sul01}, and \citet{wh96} report similar correlations between different star formation indicators to the FIR luminosity (which is directly related to the amount of dust). \citet{hop01} analyze the effect of dust-reddening dependent on SFR applied to different SFR indicators such as UV luminosity, H$_\alpha$ flux, FIR Luminosity and radio luminosity at $1.4$~GHz. They show that discrepancies between these methods can be explained by such an effect. All these trends imply that the more luminous galaxies are dustier.

\begin{figure}
\plotone{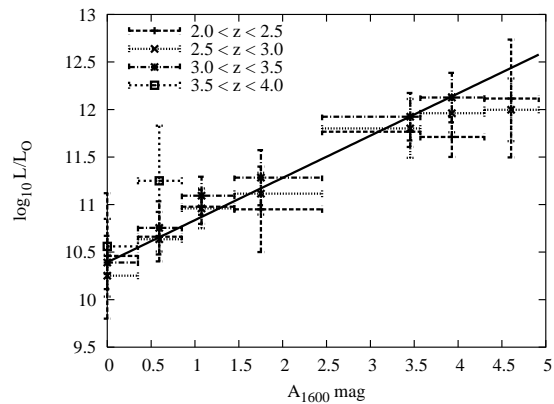}
\caption{Mean Luminosities ($L_{1600}/L_\sun$) in different attenuation ($A_{1600}$) intervals  as calculated under Case~I. The line is the mean of the least-square fits for the luminosity-attenuation correlation, for the different redshift ranges, with $\alpha=0.90\pm0.55$. \label{figal1}}
\end{figure}

In this study also we find a correlation between the intrinsic UV luminosity and the dust attenuation in these systems.  The LBGs have almost constant apparent magnitude ($\sim24.5$) but a wide range of colors. This seems to imply a correlation between the intrinsic luminosity of these galaxies and their reddening. Given our large sample and systematic analysis we can arrive at a quantitative result, but we should point out that this relation is not unique; it is dependent upon the dust geometry, dust properties and the assumed intrinsic SED. We find, 
\begin{equation}
a_{1600} \propto \left( \frac{L_{1600\ \textrm{\scriptsize{\AA}}}}{L_\sun} \right)^\alpha
\end{equation}
\noindent where $a_{1600}$ is the linear attenuation factor related to the attenuation $A_{1600}$ (in mag) as $a_{1600}=10^{0.4A_{1600}}$. The value of $\alpha$ is slightly different for case~I and case~II. For case~I, the mean rest-frame UV luminosity of all the galaxies in each attenuation interval was calculated for each redshift bin. A similar procedure was followed for case~II, except that there was no redshift binning.  Figure~\ref{figal1} is a plot of $L_{1600}/L_\sun$ vs. the attenuation at $1600$ \AA, for case~I. We find that the attenuation factor and the intrinsic UV luminosity are correlated, with $\alpha = 0.90\pm0.55$ 

This does not imply that the dust masses increase nearly linearly with UV luminosity. In a clumpy shell geometry the actual dust masses (measured by $\tau_V$) increases (from $\tau_V=1$ to $\tau_V=10$) by a factor of $10$ while the attenuation increases (from $1.8$ to $4.6$ mag) by a factor of $13.2$ ($2.8$ mag). As Figure~\ref{figAtau} shows, the $A_{1600}-\tau_V$ relation is model dependent, and the actual dust masses can be inferred from the observed attenuation for a particular geometry. In homogeneous shell geometries the attenuations would increase much more rapidly for a small increase in the dust masses.
 
As is shown in Figure~\ref{figal1}, there is no significant evolution of the attenuation-luminosity relationship with redshift; the relations in the four redshift ranges are essentially the same within the scatter of the data. This implies that the dust is mainly a direct by-product of the currently ongoing star formation. As the star-forming region ages, new stars are formed with increasing initial metallicities, reflecting the increasing metallicity of the ISM. However, the dust is mainly produced by the massive stars and SNe II and the rate of dust production or the amounts of dust produced does not evolve significantly (see sec. \ref{secSEDI} and sec. \ref{secSEDII} for a detailed discussion of the origin of the dust in LBGs). It should however be noted that because of the selection effects at the red end of the $G-\mathcal{R}$ color range, both the $3.0<z\leq3.5$ and $3.5<z\leq4$ bins are incomplete and the relationship may evolve beyond $z\sim3.0$. 

\begin{figure}
\plotone{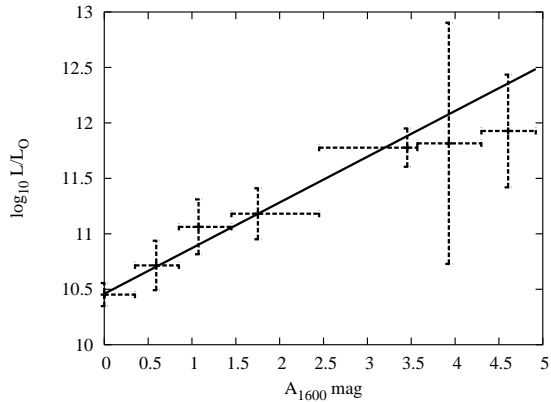}
\caption{Mean Luminosities ($L_{1600}/L_\sun$) in different attenuation ($A_{1600}$) intervals  as calculated under Case~II. The line is the least-square fit to the data, with $\alpha=0.97\pm0.49$. \label{figal2}}
\end{figure}

Figure~\ref{figal2} is the plot of $L_{1600}/L_\sun$ vs. the attenuation at $1600$ \AA, for case II. Here one SED was assumed to represent the intrinsic SED of the entire population, with a longer timescale for dust evolution. As can be seen from Figure~\ref{figsmc}(b), many more galaxies are assumed to be dust-free than in case I. This changes the dependence of the intrinsic luminosity on the dust opacity. For this case we find $\alpha=0.97\pm0.49$.

\subsection{Luminosity Weighted Attenuation Distribution Functions}
\begin{figure}
\plotone{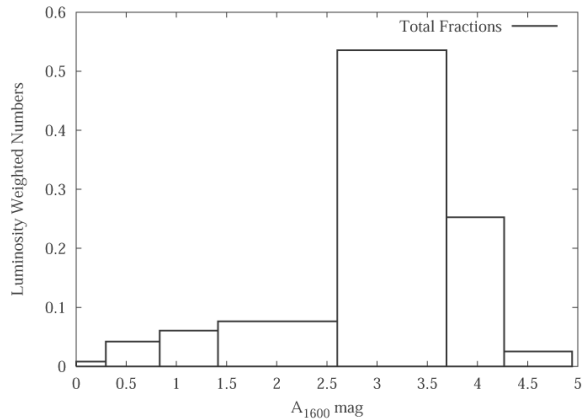}
\caption{Normalized luminosity weighted attenuation distribution function under case~I. The bulk of the galaxies are attenuated by $2.5$--$4.25$ mag (factor of $10$--$50$). \label{figadf1}}
\end{figure}

Given the correlation between dust opacity in the UV and the intrinsic UV luminosity of LBGs and the wide range of derived attenuation values, the appropriate way to present the average attenuation factor in these galaxies is a luminosity weighted average attenuation factor. A normalized luminosity weighted attenuation distribution function for these galaxies is instructive. Figure~\ref{figadf1} is the distribution function of attenuations for LBGs under case~I. As no redshift dependence was found in our analysis, the weighted attenuations for each attenuation interval were added at all redshifts. As the Figure clearly illustrates there is a large range of attenuations that these galaxies suffer, with the bulk of the luminosity attenuated by a factor of $10$ -- $50$. The assumption that all the observed galaxies are very young and are being observed during their very first star-formation episode, leads to the interpretation that all the reddening observed is a result of dust attenuation. Thus, the average value of the luminosity weighted attenuation factor for the whole sample, $18.4$ ($3.17$ mag), is considered to be an upper limit for this sample.

\begin{figure}[h]
\plotone{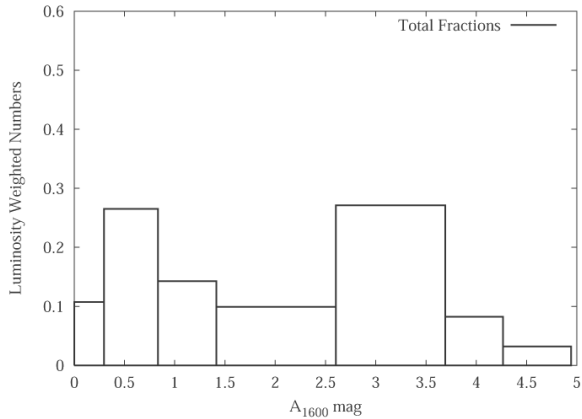}
\caption{Normalized luminosity weighted attenuation distribution function under case~II. The bulk of the galaxies are attenuated by smaller amounts. \label{figadf2}}
\end{figure}

Figure~\ref{figadf2} is the distribution function of attenuations under case~II. This sample, also, exhibits a large range in attenuations, with the bulk of the luminosity however suffering lower attenuations and only a small population being heavily attenuated. The average value of the luminosity weighted attenuation factor under this case is $5.7$ ($1.88$ mag). This value can be understood as a lower limit of the attenuation factor for the sample as a whole. A comparison of these two figures clearly illustrates the differences which are a result of the different initial assumptions. 

\section{DISCUSSION}
\subsection{The UV Attenuation Factor and its Implications for the Star Formation Rate History of the Universe}
 Our results have significant implications for the star formation history of the universe. UV luminosity in star-forming galaxies relates directly to the number of short-lived, high-mass O \& B stars, and this gives, in principle, a measure of the actual SFR after appropriate correction for dust attenuation. Therefore, the estimates of SFR at high-$z$ are closely related to the FUV attenuation factor. 

In recent years many estimates for the FUV attenuation factor at high-$z$ have been made based upon different methods. Based on metallicity arguments \citet{mad96} estimated the attenuation factor to be zero. Very soon however it was recognized that the UV luminosity is substantially suppressed by dust, and using the Calzetti law to estimate the attenuation factor from the observed UV-slope, many have found values $\sim 4-15$ \citep{meu97,mhc99,ste99,mas01b}. All these authors exploit the locally calibrated relationship between the ratio of FIR and UV fluxes and the UV spectral slope $\beta$. As shown in section~\ref{secDAf} this law can be reproduced by models for galaxies with intermediate optical depths, but there is no reason to believe that all the galaxies have similar UV-slopes or an unique value of the dust optical depth, given the wide range of colors they exhibit. \citet{pet98} obtained an attenuation factor of $2.5 - 6.3$ for a sample of 5 LBGs, by estimating SFR using H$_\beta$ luminosity and requiring that SFR$^\prime_{H\beta}=$SFR$^\prime_{UV}$, where SFR$^\prime$ are the SFRs corrected for attenuation. Estimates of the SFR derived from the H$_\beta$ luminosity are corrected for dust attenuation based upon the change in the Balmer decrement. These lines are produced in the ionized envelopes of O Stars and not B stars. Furthermore the stellar continuum, which would be measured in SFR$^\prime_{UV}$, suffers only about half the reddening suffered by the ionized gas \citep{cal01}, which makes this method unreliable. \citet{nan02} and \citet{sei02} use the X-ray luminosity to derive UV attenuations, using the assumption that the locally derived correlations between X-ray luminosity and bolometric luminosity holds for all redshifts. This is unlikely because the sources of X-ray luminosity evolve through time; in the local universe the X-ray luminosity is dominated by low mass X-ray binaries whereas at large redshifts supernova remnants of massive stars are likely to be the dominant X-ray sources. See Table~\ref{tabAttn} for different values for the UV attenuation obtained in the recent years. 

\begin{deluxetable}{lccc}
\tablecaption{Different values for the UV attenuation factors obtained in recent years\label{tabAttn}}
\tablehead{
\colhead{Method} & \colhead{Sample size} & \colhead{UV attenuation factor} & \colhead{Reference}
}
\startdata
Metallicity&$60$&$0$&1\\
Calzetti law&$60$&$3-15$&2\\
H$_\beta$ luminosity&$5$&$2.5-6.3$&3\\
Calzetti law&$25$&$5.4$&4\\
UV slope&$48$&$4.8$&5\\
Multi-wavelength analysis&\nodata&$8$&6\\
O III emission&$5$&$3$&7\\
Calzetti law&$1067$&$12\pm2$&8\\
X-Ray luminosity&$24$&$4$&9\\
X-ray luminosity&\nodata&$6.31$&10\\
UV slope&$906$&$5.7-18$&{\bf This Work}\\
\enddata
\tablerefs{(1) Madau et al. 1996; (2)Meurer et al. 1997;  (3) Pettini et al. 1998; (4) Meurer, Heckman, \& Calzetti 1999; (5) Steidel et al. 1999; (6) Adelberger \& Steidel 2000; (7) Teplitz et al. 2000; (8) Massarotti, Iovino, \& Buzzoni 2001b; (9) Seibert, Heckman, \& Meurer 2002; (10) Nandra et al. 2002}

\end{deluxetable}

Most of these methods arrive at the UV attenuation from observations at different wavelength regimes and then use the Calzetti attenuation curve to arrive at the UV attenuation values. Use of a single attenuation curve for all the galaxies at all redshifts is not valid. Moreover, estimates based on small samples probably do not represent the whole. Working with larger samples has given rise to a large range in these estimates even by these methods \citep{meu97,mas01b}. Also, these authors quote simple average values of the attenuations whereas we feel that luminosity-weighted attenuation values are more valid as the star formation in these high-$z$ systems occurs in galaxies with a wide range of luminosities and is apparently dominated by star-formation in extremely luminous systems.

\begin{figure}
\plotone{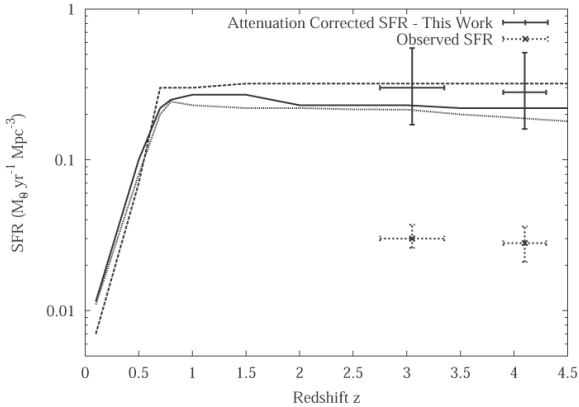}
\caption{Star formation rate history of the universe. The lines are from \citet{c&e01}, the solid line is for models that allow for luminosity evolution, the dashed line (dotted in the electronic version) is the one for density evolution and the dotted line (dashed line in the electronic version) allows for both luminosity and the density evolution. The crosses with the dotted error bars (colored blue in the electronic version) are the observed SFRs \citep{ste99}, the solid error bars (colored red in the electronic version) are our attenuation corrected values, with the lower limit obtained by applying the Case~II correction of a factor of $5.7$ and the upper limit obtained by applying the Case~I correction factor of $18.4$. \label{figmadplot}}
\end{figure}

Quite encouragingly, our results of dust attenuation correction of the UV data is consistent with the constraints imposed by the cosmic infrared background (CIRB).  \citet{c&e01} use a variety of evolutionary models to fit the local mid-IR luminosity function of galaxies with the CIRB. They deduce that the dust enshrouded SFR density peaks at $z=0.8$ and remains flat up to a redshift of $z\sim4$. In Figure~\ref{figmadplot} the three curves are the SFRs predicted by three FIR models \citep{c&e01} compared to the observed SFR\footnote{Cosmology used to estimate SFR: $\Omega_M=0.3,\ \Omega_\Lambda=0.7,\ H_0=50\ \textrm{km}\ \textrm{s}^{-1}\ \textrm{Mpc}^{-1}$} \citep{ste99} and the attenuation corrected values. The lower limit on our points is obtained by applying our lower attenuation limit ($5.7$) to the SFR and the upper one by applying our upper attenuation limit of $18.6$ to the observed SFR. If the upper limit applies, it implies that the visual/UV surveys detect all the star-formation at this redshift, but, if the lower limit applies, it indicates a population of dust enshrouded galaxies which are not seen.

Low-luminosity galaxies at high-$z$ which may or may not be very dusty are probably missed by current UV/optical surveys. Hence they would also be missed in the correlation between the attenuation and luminosity of the systems at high-$z$. But if the dust formation in high-$z$ systems is related to the evolution of massive stars and not of low-mass stars as it is currently in the Milky Way, then these galaxies will be intrinsically very luminous and be probably be detected by the UV/optical surveys.

\section{CONCLUSIONS}
We have presented our analysis of the dust in Lyman break galaxies at $z\sim3$. Our sample consisted of $906$ sources with spectroscopic redshifts, not identified as AGN or QSOs. Our main conclusions are as follows:
\begin{enumerate}
\item The wide range of the $G-\mathcal{R}$ colors of the LBGs, with most of them having similar apparent magnitudes suggests a major influence of dust reddening. Possible variation in chronological ages of the star-forming episode seems to be insufficient to account for this wide range of colors. 
\item We have used radiative transfer models incorporating SMC-like dust in a shell geometry (consistent with the Calzetti law) which also includes the dependence of the attenuation function on the dust optical depths. The LBGs can be represented by an inner dust free region with high SFRs surrounded by a shell of clumpy dust.
\item The observed $G-\mathcal{R}$ colors appear to be incompatible with the presence of MW-like dust in these systems. The dust in LBGs is likely to be SMC-like, as found in other star-bursting systems.
\item Complete resolution of the age-reddening degeneracy is not possible with a single observed rest-frame UV-color. However, the dust attenuation can be constrained by considering limiting cases where dust formation follows evolutionary timescales of massive stars (case~I) and where the dust timescales follow evolution of intermediate mass ($2.5-3\ M_\sun$) stars (case~II). The sources of dust production in case~I are most likely to be SNe II. In case~II the dust is most likely to be produced during the pre-planetary nebula stages of AGB stars.
\item We performed an independent test of our reddening/attenuation models with limited set of LBGs for which $V-H$ colors are available. This test points toward an identical range of UV attenuations as found with the larger $G-\mathcal{R}$ color set.
\item Case~I requires IMFs with upper mass limits increasing up to $200\ M_\sun$ at $3.5\leq z < 4.0$ and initial metallicity $Z=0.01\ Z_\sun$ in order to match the bluest observed colors found among the LBGs.
\item Both limiting cases I and II result in attenuation factors which exhibit a pronounced dependence on the luminosities of the LBGs:
\begin{displaymath}
a_{1600}\textrm{ (Case~I)} \propto (L_{1600\ \textrm{\scriptsize{\AA}}}/L_\sun)^{0.90\pm0.55}
\end{displaymath}
\begin{displaymath}
a_{1600}\textrm{ (Case~II)} \propto (L_{1600\ \textrm{\scriptsize{\AA}}}/L_\sun)^{0.97\pm0.49}
\end{displaymath}
\item We do not find any $z$-dependence in the dust properties of the LBGs up to $z\sim3.5$, i.e. galaxies of similar luminosity are attenuated by similar amounts over the range $2\leq z<4$. This suggests that the dust present in these systems is a direct result of the currently ongoing star formation episode.
\item The LBGs suffer a wide range of attenuations from zero to five mag; examined under both the limiting cases.
\item Our luminosity  weighted average attenuation factors for the rest-frame luminosity at $1600$~\AA$\ $ for the LBG sample are $18.5$ (case~I) and $5.7$ (case~II).
\item When applied to the observed star formation rates at $z\sim3$ and $z\sim4$, the attenuation corrected star formation rates are consistent with model predictions based upon galaxy counts and the cosmic IR background. This implies that the bulk of the star formation during the epoch $2<z<4$ is directly detectable by current optical surveys, but requires substantial correction for internal dust attenuation. 
\end{enumerate}

\acknowledgments
We thank Charles Steidel for providing us with the photometric data on the LBGs. We would also like to thank Eric Feigelson for sharing his thoughts on the dust productions in the early universe with us. The first author would like to thank Daniele Pierini for stimulating and helpful discussions throughout most of this work. We are grateful to the anonymous referee for a number of helpful and constructive comments and suggestions, which have resulted in improvements in this paper. This work made use of the NASA Astrophysical Data System (ADS). This research was supported by the NASA grants NAG5-9376 and NAG5-9202, which we acknowledge with gratitude.

\end{document}